\documentclass[amsmath,amssymb,twocolumn]{revtex4}
\usepackage{graphicx}
\usepackage{amscd,amsmath,amsthm,amsfonts,amssymb}
\def\PT{$\cal{PT}$}
\def\[{\begin{equation}}
\def\]{\end{equation}}
\advance\textheight -16mm

\begin{document}
\title{Nonlinear behaviors of parity-time-symmetric lasers}
\author{Jianke Yang}
\address{Department of Mathematics and Statistics, University of Vermont, Burlington, VT 05401, USA}

\begin{abstract}
We propose a time-dependent partial differential equation model to investigate the dynamical behavior of the parity-time (\PT) symmetric laser during the nonlinear stage of its operation. This model incorporates physical effects such as the refractive index distribution, dispersion, material loss, nonlinear gain saturation and self-phase modulation. We show that when the loss is weak, multiple stable steady states and time-periodic states of light exist above the lasing threshold, rendering the laser multi-mode. However, when the loss is strong, only a single stable steady state of broken \PT symmetry exists for a wide range of the gain amplitude, rendering the laser single-mode. These results reveal the important role the loss plays in maintaining the single-mode operation of \PT lasers.
\end{abstract}

\maketitle

Parity-time (\PT) symmetry was first introduced as a non-Hermitian generalization of quantum mechanics in 1998, where it was reported that a class of complex potentials possessing the \PT symmetry could also feature all-real spectra
\cite{Bender1998}. This concept later spread to optics, where a judicious balancing of gain and loss constitutes a \PT-symmetric system \cite{Musslimani2008,Segev2010,coupler1,PT_lattice_exp_2012,Peng2014}. Properties of \PT systems have been extensively studied in the past ten years (see \cite{Kivshar_review, Yang_review} for reviews). More importantly, applications of \PT symmetry have started to emerge \cite{Longhi_2010, Feng2013,Zhang2014,Mercedeh2014}. Of particular interest is a \PT microring laser \cite{Zhang2014,Mercedeh2014}. By intentionally introducing loss into the laser cavity, it was shown that these lasers are capable of single-mode operations. These operations were explained using a linear coupled-mode ordinary differential equation (ODE) model. However, it is well known that lasing is an inherent nonlinear process. In order to investigate the nonlinear stage of these \PT lasers, certain nonlinear coupled-mode ODE models were used \cite{Demetri_2015,Ge_2016}. But such ODE models did not account for the effects of refractive-index distributions, dispersion, nonlinear self-phase modulation, and sometimes nonlinear modal interactions. In \cite{Ge_2016}, a steady-state \emph{ab initio} laser theory was also employed, but a dynamic (time-dependent) model would be more desirable since only a dynamical model could address the question of stability of steady-state laser states.

In this paper, we propose a qualitative time-dependent partial differential equation (PDE) model to investigate the dynamical behavior of \PT lasers during their nonlinear stage of operation. This model incorporates the physical effects such as the refractive index distribution, dispersion, spatially-modulated linear loss, spatially-modulated nonlinear saturable gain, and nonlinear self-phase modulation. Even though this PDE model is more elaborate than the previous coupled-mode ODE models, it is still simple enough for theoretical analysis. We apply this model to a \PT-laser configuration similar to that in \cite{Mercedeh2014} and show that, when the loss is weak, multiple stable steady states and time-periodic states of light exist above the lasing threshold, rendering the laser multi-mode. However, when the loss is strong, only a single stable steady state of broken \PT symmetry exists for a wide range of the gain amplitude, rendering the laser single-mode. These results show that the loss plays a critical role in maintaining the single-mode operation of \PT lasers.

The qualitative PDE model we propose for \PT lasers is
\[  \label{e:PDE}
i\Psi_t+\Psi_{xx}+V(x)\Psi+\sigma |\Psi|^2\Psi-i \frac{G(x)}{1+|\Psi|^2}\Psi=0,
\]
where $\Psi$ is a complex envelope function of the light's electromagnetic field, $t$ and $x$ are the time and space variables,
$V(x)=n(x)+i\Gamma(x)$ is the complex potential whose real part $n(x)$ describes the refractive-index distribution and the imaginary part $\Gamma(x)$ represents the linear (material) loss, $\sigma$ is the coefficient of nonlinear self-phase modulation, $G(x)$ is the spatially-modulated linear gain, which is saturable at high intensities. All variables have been normalized. This model is one-dimensional for simplicity, but its higher-dimensional extension is straightforward.

\begin{figure}[tbh!]
\includegraphics[width=0.48\textwidth]{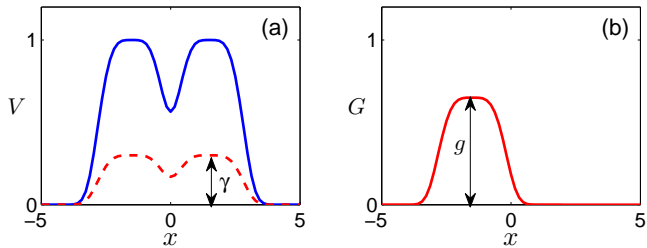}

\smallskip
\caption{(a) Profiles of the refractive index distribution $n(x)$ (solid blue) and the linear loss function $\Gamma(x)$ (dashed red). (b) Profile of the gain function $G(x)$.}
\label{f:fig1}
\end{figure}

We apply this model to the \PT-laser configuration in \cite{Mercedeh2014}, where two identical microrings are placed adjacent to each other, and the gain is applied to only one of them. Corresponding to this configuration, we
consider two one-dimensional regions with identical refractive-index and loss distributions, but the gain is applied only to the left region. Thus, we choose functions in the model (\ref{e:PDE}) as
\[
n(x)=n_0 \;[f_1(x+x_0)+f_1(x-x_0)],
\]
\[
\Gamma(x)=\gamma\; [f_2(x+x_0)+f_2(x-x_0)],
\]
\[
G(x)=g \hspace{0.04cm} f_3(x+x_0),
\]
where $f_1(x), f_2(x), f_3(x)$ are spatial distributions of the refractive index, linear loss and gain in each region, $n_0, \gamma$ and $g$ are their peak values, and $-x_0$ and $x_0$ are the center positions of the two regions. For simplicity, we select $f_1, f_2, f_3$ to be the same super-Gaussian function
\[
f_1(x)=f_2(x)=f_3(x)=e^{-x^4/4}.
\]
In addition, we select $n_0=1$ and $x_0=1.5$. Regarding the nonlinear coefficient $\sigma$, we set $\sigma=0.5$, which is self-focusing nonlinearity. Then, the remaining free parameters in our model are the loss coefficient $\gamma$ and the gain amplitude $g$. The resulting profiles of $n(x)$, $\Gamma(x)$ and $G(x)$ are displayed in Fig.~\ref{f:fig1}. When $g=2\gamma$, the effective linear potential $V(x)-iG(x)$ is \PT-symmetric.

Now, we investigate solution behaviors of the model (\ref{e:PDE}) for a fixed loss coefficient $\gamma$ and a tunable gain amplitude $g$. Steady-state (soliton) solutions are of the form
\[
\Psi(x, t)=e^{i\mu t}\psi(x),
\]
where $\psi(x)$ is a localized function solving
\[ \label{e:soliton}
\psi_{xx}+V(x)\psi+\sigma |\psi|^2\psi-i \frac{G(x)}{1+|\psi|^2}\psi=\mu \hspace{0.05cm} \psi,
\]
and $\mu$ is a real frequency parameter. Unlike in \PT-symmetric nonlinear systems \cite{Yang_review}, solitons in Eq. (\ref{e:soliton}) exist only at isolated frequency values due to the saturable gain, which breaks the \PT symmetry even if the effective linear potential $V(x)-iG(x)$ is \PT-symmetric. We will compute these isolated solitons by the squared operator method developed in \cite{Yang_SOM}, which yields the soliton profile $\psi(x)$ as well as the frequency value $\mu$ simultaneously. Linear stability of these solitons can be determined by computing the spectrum of the linear-stability operator of these solitons by the Fourier-collocation method \cite{Yang_SIAM}.

\begin{figure}[tb!]
\includegraphics[width=0.48\textwidth]{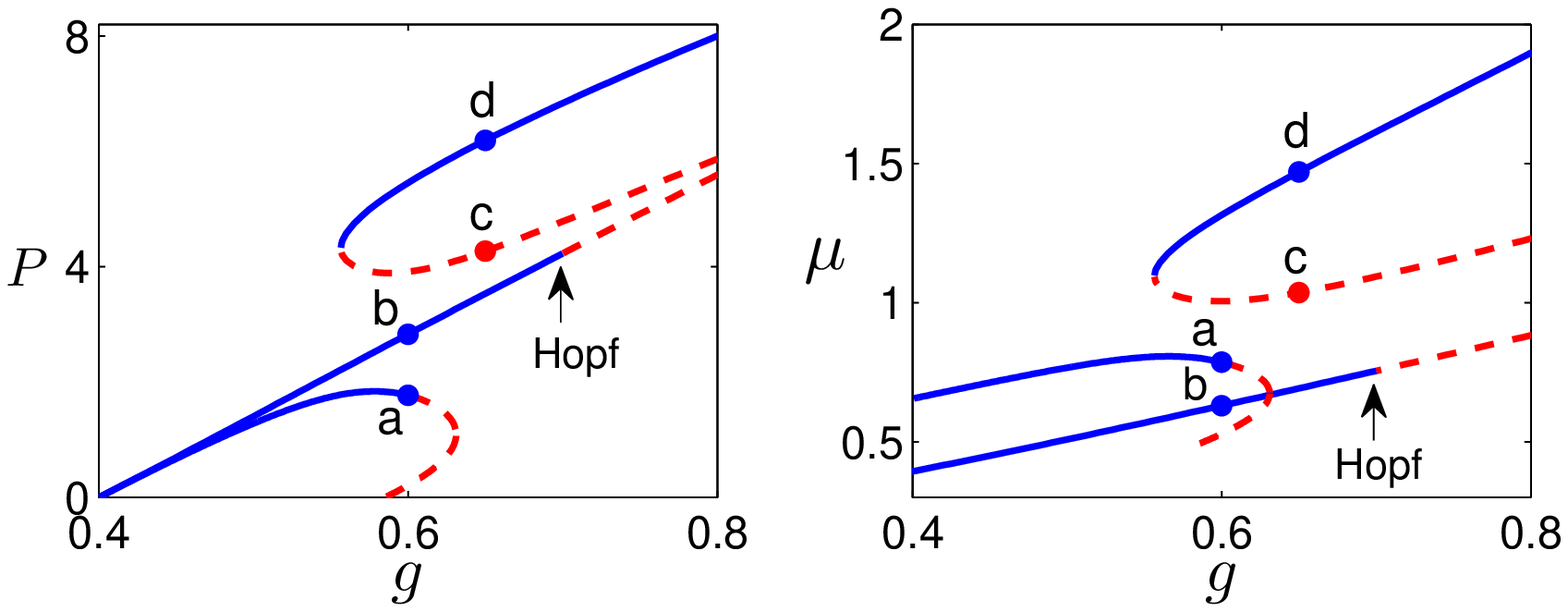}

\vspace{0.1cm}
\includegraphics[width=0.48\textwidth]{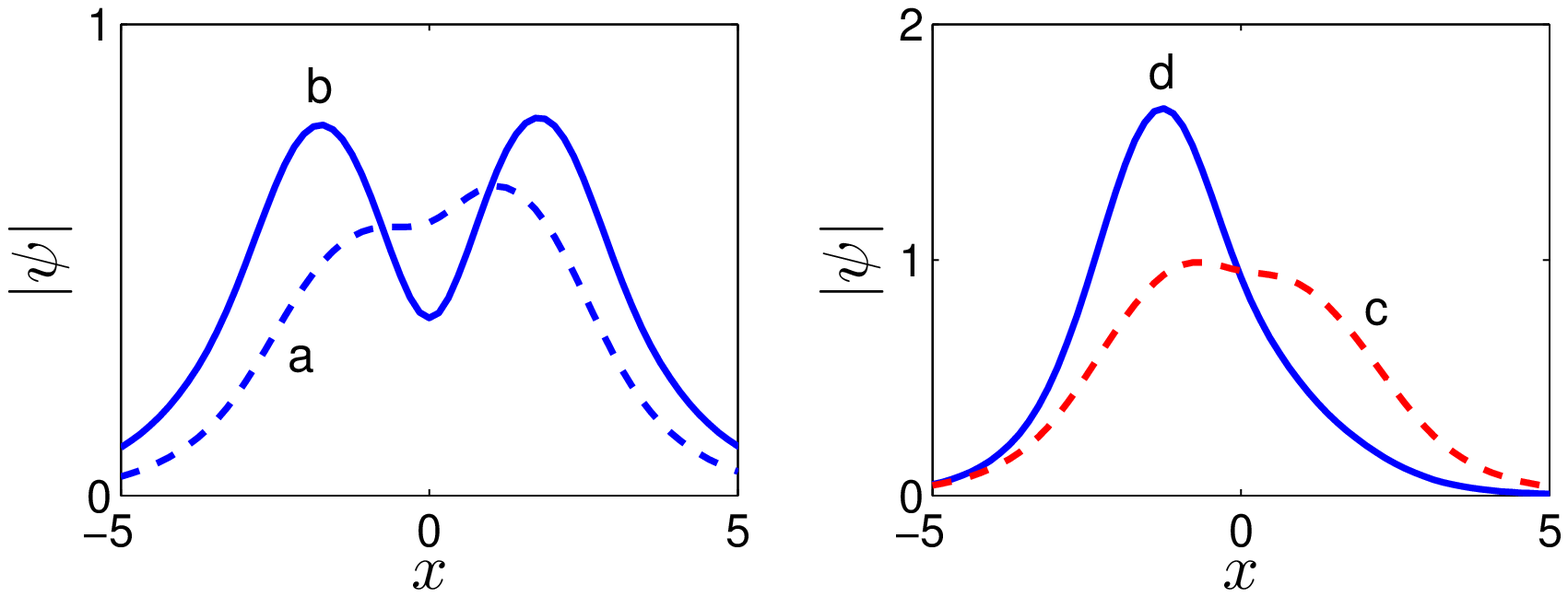}
\smallskip
\caption{Soliton branches versus the gain coefficient $g$ at the lower-loss value of $\gamma=0.2$. Upper row: power and frequency branches (solid blue indicates stable solitons, and dashed red indicates unstable solitons).
Lower row: soliton profiles at the marked points of the power/frequency branches. } \label{f:fig2}
\end{figure}

First we consider the lower-loss case, where we set $\gamma=0.2$. In this case, lasing occurs (i.e., infinitesimal light starts to amplify) when $g>2\gamma=0.4$. Thus, at the lasing threshold $g=2\gamma=0.4$, the linear system is in \PT-symmetric state.
Above this lasing threshold, we have found a number of soliton branches, which are displayed in Fig.~\ref{f:fig2}. Two of these branches bifurcate off the zero amplitude at $g=0.4$. On the higher-power branch, solitons have an approximately symmetric profile (see `b' on the lower panel) and can be said to be in \PT-symmetric state; while on the lower-power branch, solitons have a slightly asymmetric profile (see `a' on the lower panel). At $g$ increases, the `a' branch undergoes a fold bifurcation and disappears, while the `b' branch persists. When $g>0.56$, two additional soliton branches (the `c,d' branches) appear through another fold bifurcation, and their powers have a non-zero minimum threshold. On their higher-power branch, solitons reside primarily in the gain region (see `d' on the lower panel) and can be said to be in broken-\PT-symmetry state \cite{Demetri_2015,Ge_2016}; while on the lower-power branch, solitons reside in both the gain and loss regions (see `c' on the lower panel).

Linear stability of the soliton branches in Fig.~\ref{f:fig2} is also marked in the same figure. It is seen that the two soliton branches bifurcating from $g=0.4$ are stable when $g$ is close to 0.4, but lose stability when $g$ is larger (the `a' branch loses stability when $g>0.60$, while the `b' branch loses stability when $g>0.70$). Both losses of stability are due to Hopf bifurcations, where a pair of complex linear-stability eigenvalues cross the imaginary axis. After the Hopf bifurcation, the solitons become unstable. Simultaneously, stable time-periodic bound states appear.

\begin{figure}[tb!]
\includegraphics[width=0.48\textwidth]{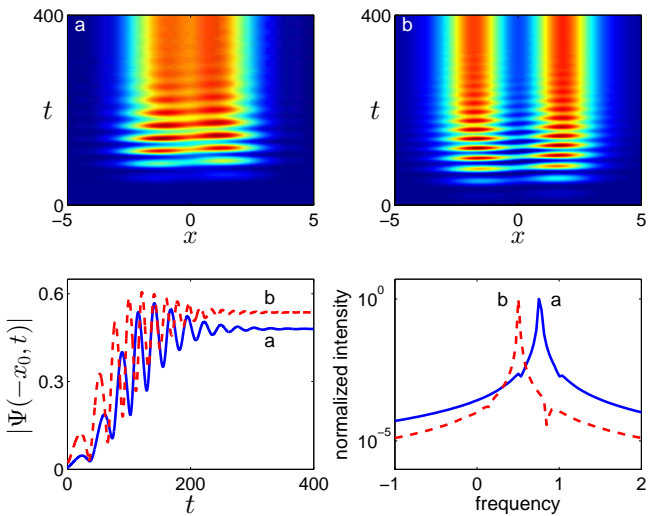}

\smallskip
\caption{Evolutions of Eq. (\ref{e:PDE}) for two different infinitesimal random-noise initial conditions when $\gamma=0.2$ and $g=0.5$. The lower panels show the amplitude evolutions and frequency spectra at the gain center $x=-1.5$.}
\label{f:fig3}
\end{figure}

Now we examine how light behaves in this lower-loss case. For this purpose, we consider two gain amplitudes, $g=0.5$ and $0.72$. When $g=0.5$, Fig.~\ref{f:fig2} shows that there are two solitons which are both stable. Numerically, we have found that any infinitesimal initial condition would evolve toward one of these two stable solitons. To illustrate, numerical simulations of Eq. (\ref{e:PDE}) for
two different infinitesimal random-noise initial conditions are displayed in Fig.~\ref{f:fig3}. The upper panels show that the two infinitesimal initial conditions are attracted toward different solitons. Amplitude evolutions and frequency spectra at the gain center $x=-x_0$, shown in the lower panels, confirm that the final states are solitons on the `a' and `b' branches of Fig.~\ref{f:fig2} respectively. Since these two solitons have different frequencies, this laser cavity can produce light of different frequencies and is thus not a single-mode laser.

\begin{figure}[tb!]
\includegraphics[width=0.48\textwidth]{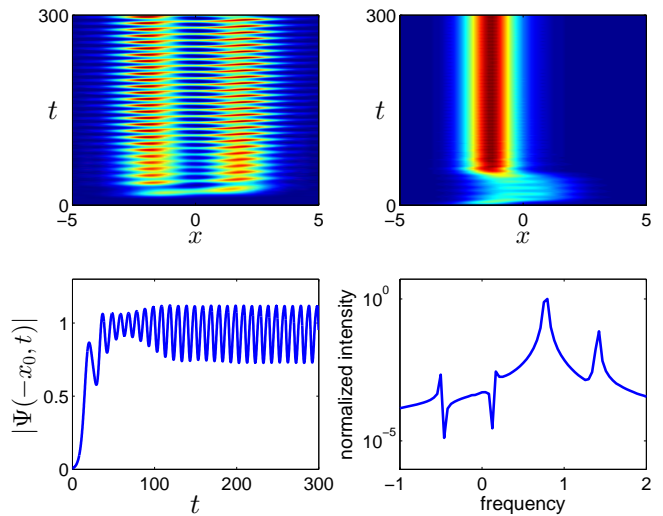}

\smallskip
\caption{Evolutions of Eq. (\ref{e:PDE}) for two different initial conditions when $\gamma=0.2$ and $g=0.72$ (the initial condition is infinitesimal in the upper left panel and has finite amplitude in the upper right panel).
The lower panels show the amplitude evolution and frequency spectrum at the gain center $x=-1.5$ for the upper left panel.}
\label{f:fig4}
\end{figure}

At the higher gain amplitude $g=0.72$ (which is above the Hopf bifurcation point on the `b' branch of Fig.~\ref{f:fig2}), the only stable soliton is on the `d' branch. But a stable time-periodic bound state also exists due to the Hopf bifurcation.
In this case, we have found numerically that all infinitesimal initial conditions are attracted toward that time-periodic bound state, see Fig.~\ref{f:fig4} (upper left panel). This temporal periodicity indicates that the output of light has multiple frequencies, as is evidenced in the lower right panel of Fig.~\ref{f:fig4}. Thus, this laser is not single-mode either. For some finite-amplitude initial conditions, though, light can evolve toward the stable soliton on the `d' branch, see the upper right panel of Fig.~\ref{f:fig4}.

\begin{figure}[tb!]
\includegraphics[width=0.48\textwidth]{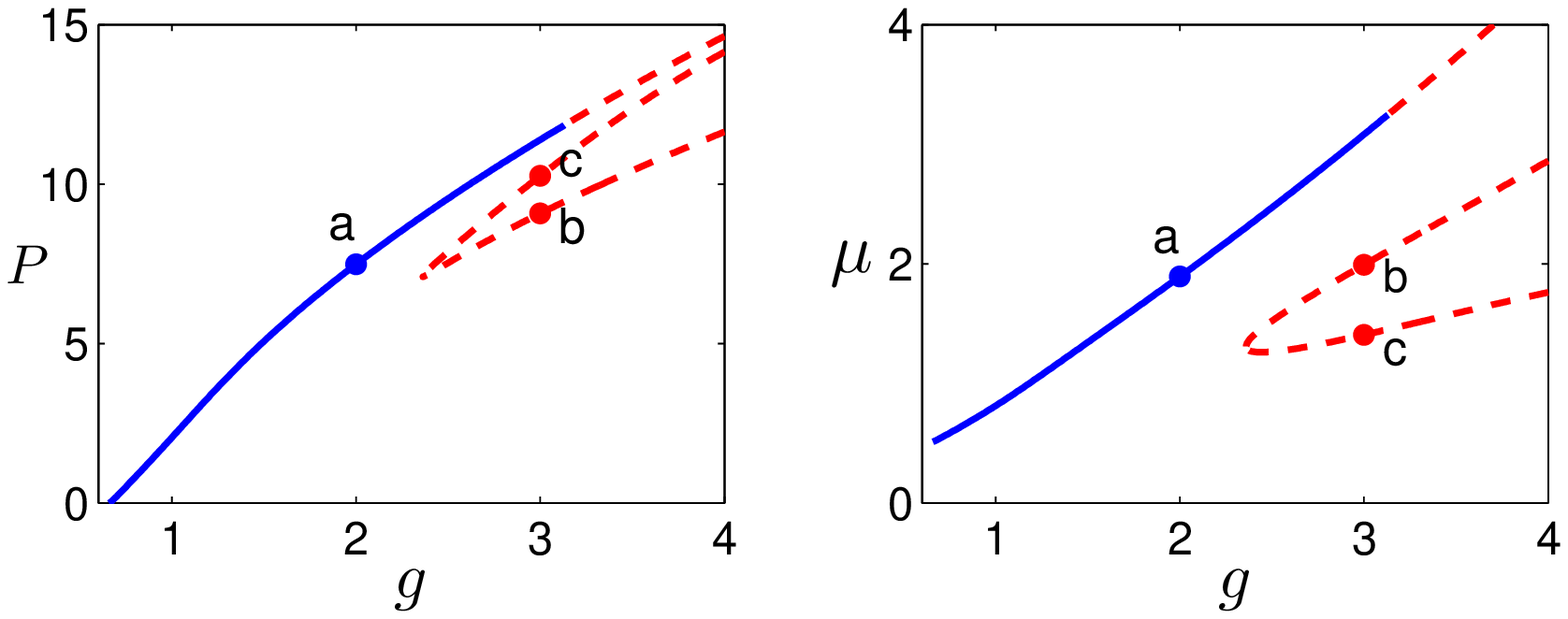}

\vspace{0.1cm}
\includegraphics[width=0.48\textwidth]{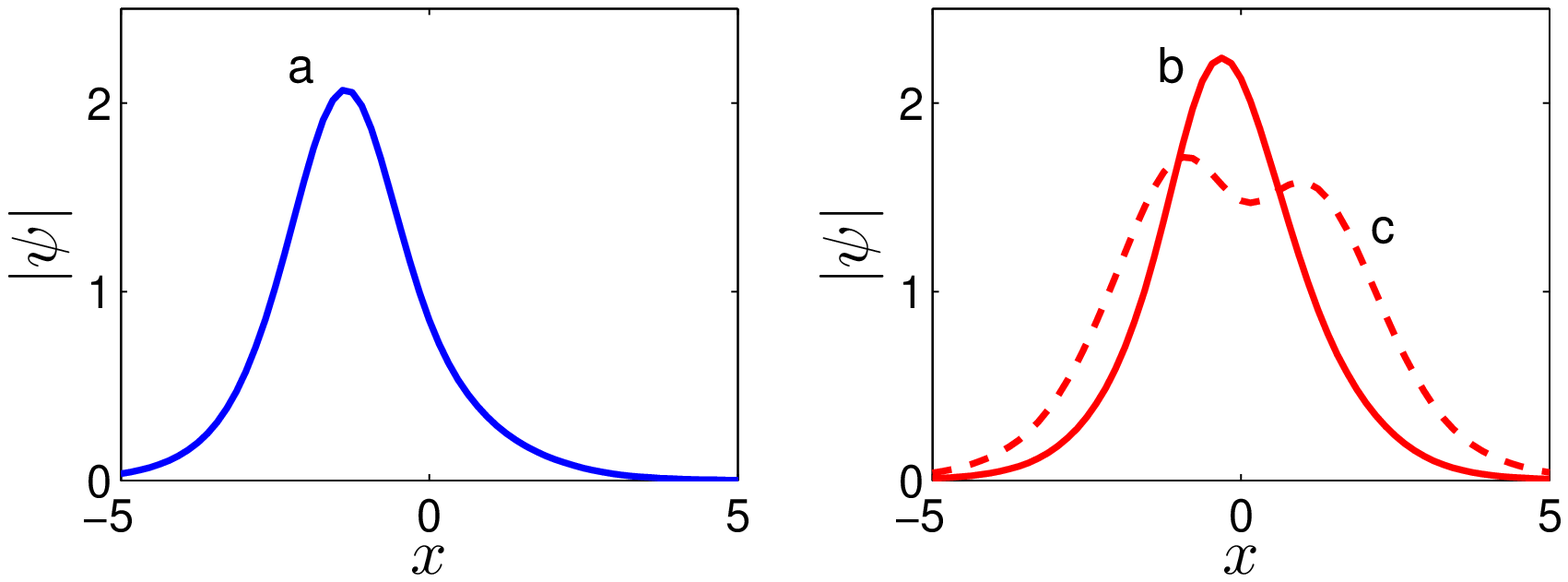}
\smallskip
\caption{Soliton branches versus the gain coefficient $g$ at the higher-loss value of $\gamma=0.5$. Upper row: power and frequency branches (solid blue for stable solitons and dashed red for unstable solitons).
Lower row: soliton profiles at the marked points of the power/frequency branches.  } \label{f:fig5}
\end{figure}

\begin{figure}[tb!]
\includegraphics[width=0.48\textwidth]{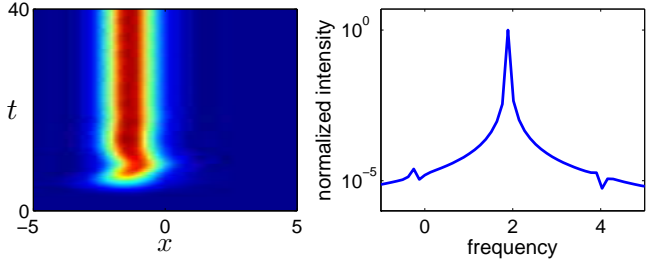}

\smallskip
\caption{Evolution of Eq. (\ref{e:PDE}) for an infinitesimal random-noise initial condition with $\gamma=0.5$ and $g=2$. The right panel shows the frequency spectrum at the gain center $x=-1.5$. }\label{f:fig6}
\end{figure}

Next we consider the higher-loss case, where we set $\gamma=0.5$. In this case, lasing occurs when $g>0.66$. At this lasing threshold, the linear system is in broken \PT-symmetry state. Above this threshold, soliton branches versus the gain amplitude $g$ are displayed in Fig.~\ref{f:fig5}. It is seen that a branch of stable solitons bifurcates from the zero amplitude at the lasing threshold, and it loses stability when $g>3.1$. These solitons reside primarily in the gain region and thus have broken \PT-symmetry (see the lower left panel). Meanwhile, two branches of unstable solitons appear through a fold bifurcation when $g>2.36$, and these solitons reside in both the gain and loss regions (see the lower right panel). The striking feature in this case is that, over a wide gain interval of $0.66<g<3.1$, there is a single stable soliton, and no other stable coherent states (such as time-periodic states) exist. Our numerical simulations show that on this wide gain interval, all infinitesimal initial conditions evolve toward this single stable soliton (see Fig.~\ref{f:fig6}), thus the laser is in single-mode operation.

By comparing the lower-loss operation in Figs. \ref{f:fig2}-\ref{f:fig4} and higher-loss operation in Figs. \ref{f:fig5}-\ref{f:fig6}, we see that when the loss is weak, multiple stable solitons and time-periodic bound states exist above the lasing threshold, rendering the laser multi-mode. However, a strong loss can eliminate those multiple stable solitons and time-periodic states, leaving the system with a single stable soliton of broken \PT symmetry and thus rendering the laser single-mode. These results show that the loss is instrumental in maintaining the single-mode operation of \PT lasers.

Some of the solution behaviors in the PDE model for the configuration of Fig.~1 can be understood from a simpler ODE model,
\[\frac{d\Phi_1}{dt}=-\alpha \Phi_1 + \frac{\beta}{1+|\Phi_1|^2}\Phi_1 +i\Phi_2,   \label{e:ODE1}\]
\[\frac{d\Phi_2}{dt}=-\alpha \Phi_2 +i\Phi_1,   \hspace{2.2cm}   \label{e:ODE2} \]
where $\Phi_1$ and $\Phi_2$ are the amplitudes of the supermodes in the gain (left) and loss (right) regions of the laser cavity respectively, $\alpha$ is the linear material loss in both regions, and $\beta$ is the linear growth rate of the saturable gain. The parameter values are normalized with respect to the coupling constant between the gain and loss regions. Similar but different ODE models have been used in \cite{Demetri_2015,Ge_2016} to study various \PT-laser configurations.

\begin{figure}[tb!]
\includegraphics[width=0.48\textwidth]{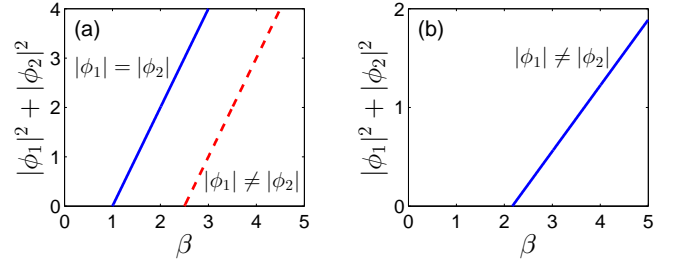}

\smallskip
\caption{Power curves of steady states versus gain parameter $\beta$ in the ODE model. (a) $\alpha=0.5$ (lower loss); (b) $\alpha=1.5$ (higher loss). Solid blue: stable states; dashed red: unstable states. } \label{f:fig7}
\end{figure}

Steady states of the above ODE model are of the form $\Phi_{1, 2}(t)=e^{i\omega t}\phi_{1, 2}$,
where $\omega$ is a real frequency parameter, and $\phi_1, \phi_2$ are amplitude constants. Substituting this steady state into the ODE model, we find the following solutions,
\[  \label{e:s1}
(1)\; \omega=\pm \sqrt{1-\alpha^2}, \hspace{0.2cm} |\phi_1|=|\phi_2|=\sqrt{\beta/(2\alpha)-1}; \hspace{0.68cm}
\]
\[  \label{e:s2}
(2) \; \omega=0, \hspace{0.2cm} |\phi_1|=\sqrt{\beta/(\alpha+1/\alpha)-1}, \hspace{0.2cm} |\phi_2|=|\phi_1|/\alpha.
\]
The two solutions (\ref{e:s1}) exist only when the loss is weak ($\alpha<1$). They have identical amplitudes but different frequencies. In each solution, the modal amplitudes in the gain and loss regions are the same ($|\phi_1|=|\phi_2|$), and are thus in \PT-symmetric state. They bifurcate out from the zero amplitude at the lasing threshold $\beta=2\alpha$, where the linear counterpart of the ODE model is \PT-symmetric. The other solution (\ref{e:s2}) exists for all loss values of $\alpha$. In this solution, the modal amplitudes in the gain and loss regions are different, and are thus in broken-\PT-symmetry state. This solution
bifurcates out from the zero amplitude at $\beta=\alpha+1/\alpha$. To illustrate these solutions, we choose two loss values of $\alpha=0.5$ (lower loss) and $\alpha=1.5$ (higher loss), where the power values $|\phi_1|^2+|\phi_2|^2$ of these solution branches versus the gain parameter $\beta$ are displayed in Fig.~\ref{f:fig7}(a,b) respectively. Linear stability of these steady states has also been determined and marked on the figure. At lower loss, the two \PT-symmetric solutions (\ref{e:s1}) are both stable, rendering the laser multi-mode; and the broken-\PT-symmetry solution is unstable (see Fig.~\ref{f:fig7}a). Our numerics shows that these two stable solutions (\ref{e:s1}) attract all infinitesimal initial conditions. At higher loss (see Fig.~\ref{f:fig7}b), the sole broken-\PT-symmetry solution (\ref{e:s2}) is stable and attracts all infinitesimal initial conditions.

When comparing these ODE-model results with those PDE ones, we can see that at lower loss, the solid blue branch of two \PT-symmetric steady states in Fig.~\ref{f:fig7}(a) is the counterpart of the two solid blue branches bifurcating from $g=0.4$ in Fig.~\ref{f:fig2} (upper left panel), and the dashed red branch of broken-\PT-symmetry states in Fig.~\ref{f:fig7}(a) is the counterpart of the dashed red branch bifurcating from $g\approx 0.58$ in Fig.~\ref{f:fig2}. At higher loss, the solid blue branch of broken-\PT-symmetry states in Fig.~\ref{f:fig7}(b) is the counterpart of the solid blue branch bifurcating from the zero power in Fig.~\ref{f:fig5}. Notice that at lower powers, solitons and their stability behaviors in the PDE model agree with their ODE counterparts. However, at higher powers, the PDE and ODE results show large differences. In particular, the ODE model fails to predict the new soliton branches from fold bifurcations at high powers, the Hopf bifurcations of soliton branches, and time-periodic bound states after Hopf bifurcations.
The reason is that this ODE model implicitly assumes a fixed supermode profile in each of the gain and loss regions (only their amplitudes $\Phi_{1,2}$ are allowed to change with time). This assumption is reasonable for the PDE solutions at low powers, where the solutions are close to linear modes. However, at higher powers, the field profiles of PDE solutions will differ significantly from linear modes due to nonlinear self-focusing and other physical effects. In addition, the field profiles of PDE solutions can also oscillate with time (such as in time-periodic bound states, see Fig.~\ref{f:fig4}), where multiple frequencies (modes) are mixed. In such cases, the ODE model cannot be expected to give good predictions, and a dynamic PDE model will become necessary, as is done in this article.

In summary, we have proposed a time-dependent PDE model to investigate the dynamical behavior of \PT lasers during the nonlinear stage of its operation.
Our results indicate that a significant amount of material loss is important for rendering the single-mode operation over a wide range of gain values. These results shed more light on the operation of \PT lasers. In addition, we anticipate that this dynamic PDE model will help improve the functionalities of \PT lasers.

\vspace{0.2cm} \noindent This work was supported in part by Air Force Office of Scientific Research
(USAF 9550-12-1-0244) and National Science Foundation (DMS-1311730).

\end{document}